# Current-induced magnetization reversal in pseudo spin valves in current-in-plane configuration

J. Kleinlein and G. Schmidt

Institut für Physik, Martin-Luther-Universität Halle-Wittenberg, 06099 Halle, Germany

We demonstrate the current-induced complete magnetization reversal of the free layer in lateral nanowires patterned from metallic pseudo spin valve stacks. The reversal is induced by Oersted fields in conjunction with a dipolar coupling via the lateral stray fields of the two magnetic layers. Starting from the parallel state the Oersted field rotates the magnetization vectors of both magnetic layers in opposite directions so far off the wire axis that the dipolar coupling becomes strong enough to stabilize an antiparallel configuration. As simulations and experiments show this coupling is only achieved for a narrow range of layer thicknesses and for suitable wire geometries. The reversal process is demonstrated as well for straight wires as for wire segments of L-shaped wires with inhomogeneous magnetization in which domain walls are present before and after the switching process.

Current-induced magnetization switching in magnetic nanostructures has been investigated for quite some time. Experiments published so far can mainly be divided in two categories, namely transport in vertical and lateral devices, respectively. In vertical nanostructures typically consisting of nanosized spin valve structures the magnetization can be reversed [1] in zero magnetic field by spin transfer torque (STT) [2], [3] or by Oersted fields induced by a current through the structure [4]. In lateral transport, usually performed in narrow ferromagnetic wires, magnetization reversal is achieved by current-induced domain wall motion [5], [6], [7] however, the domain walls first need to be induced [8] either by locally or by globally applied magnetic fields. Moreover current-induced precessional magnetization reversal can occur in spin valve structures using ultrashort ($<1\,\text{ns}$) current pulses [9].

Here we show that in lateral nanowires patterned from giant magneto resistance [10], [11] (GMR) pseudo spin valve (PSV) layer stacks it is possible to completely reverse the magnetization of the free layer from parallel (P) to antiparallel (AP) magnetization state by a long-risetime current pulse through the wire, even if no domain wall is present before the current is applied. The reversal is caused by the Oersted fields induced by the current in combination with the dipolar coupling between the two magnetic layers of the GMR stack. In order to confirm the results micromagnetic simulations using the Mumax 2 code [12] are carried out.

The structures under investigation are tapered nanowires patterned out of a metallic multilayer. The layer stack is deposited by DC-magnetron-sputtering on a thermally oxidized silicon wafer (1 µm nominal thickness of $SiO_2$). The deposition is done in a homemade sputtering tool with a base pressure of less than $5 \cdot 10^{-9}$ mbar. The layer structure is substrate / Ta (3) / free layer (d) / Cu (5) / $Co_{50}Fe_{50}$ (10) / Ta (3) / Ru (7), units in nm. The thickness of the free layer ($Ni_{80}Fe_{20}$ unless otherwise stated) d is varied. The multilayer is covered with PMMA which is then patterned in a positive electron beam process. By evaporation and lift-off of $Al_2O_3$, a hard mask is deposited in the shape of the wire to be fabricated. By Argon ion beam etching (IBE) the pattern is transferred into the metal. All unmasked layers are removed down to the substrate. After the milling process the mask is removed by selective wet etching. The stripes are contacted by Ti (10) / Au (150) leads in a follow-up lift-off process with prior cleaning of the surface by Argon IBE to ensure good electrical contacts. Figure 1 shows SEM micrographs of model nanowires with the different contact geometries used in this study.

Figure 1(a) shows our standard geometry with large contact pads, Figure 1(b) shows a contact geometry used for 4-probe measurements and Figure 1(c) shows a L-shaped wire with 3 contacts. Unless otherwise stated all measurements are carried out at room temperature in a probe station fitted with an electromagnet that can be rotated in the sample plane by 360° and with a calibrated Hall-sensor. The resistance is measured using I/V measurements with a low bias voltage resulting in a low current through the stripe (typical order of magnitude: $50\,\mu A$). Figure 2 shows small current magneto resistance (MR) measurements of a 200 nm wide GMR nanowire with a length of $10\,\mu m$ between areal contact pads (Fig. 1 (a) shows a SEM image of the structure). The coercive fields determined from these measurements are $(1.4\pm0.5)\,mT$ for the NiFe (free) layer and $(39.6\pm1.1)\,mT$ for the CoFe (hard) electrode, respectively. In order to allow for large currents, the copper interlayer is relatively thick. Additionally the thickness of the free electrode is very small. As a consequence our nanowires show a GMR effect of less than 2 % at ambient temperature. The measurements show a sharp magnetization reversal.

We investigate the current-induced switching using a quasi static measurement technique [13] (single shot measurements). We start from a well defined parallel magnetic state by applying a large in-plane magnetic field (typically $-150\,mT$) along the stripe (reset). After reducing the magnetic field to zero ($<0.03\,mT$) the magnetizations of both electrodes remain in the parallel state and also parallel to the direction of the nanowire. The corresponding electrical resistance is low. Subsequently a single $250\,\mu s$-long current pulse of larger magnitude is applied followed by a second low bias resistance measurement. The current pulses have a rectangular shape and typical rise times in the range of a few hundred $ns$. The rise times of our current pulses increase with increasing current. This sequence is repeated with increasing current pulse magnitudes, typically up to the $mA$ range.

Figure 3(a) shows the change in resistance after a current pulse (normalized to the GMR value at zero magnetic field) of a 200 nm wide and 10 µm long stripe with a 3 nm thick NiFe electrode (standard geometry, see Fig. 1(a)) as a function of the pulse current density. Above a threshold value of $7 \cdot 10^{11}$ A/m$^2$ an increase in resistance is observed. For the highest current densities the resistance change corresponds to a complete magnetization reversal to the AP state. This reproducible switching is always reversible by applying the reset field. The current densities are calculated under the simplifying assumption that the current flows only through the copper layer. This assumption is supported by a parallel resistor model for the different layers of the stack. Using literature values of sputtered thin films for the resistances of each layer in current in plane geometry we find that about 90 % of the current flows through the Cu layer. In [14] a similar estimation is shown in more detail. Furthermore Figure 3(a) shows that after pulses with a current density close to the critical current density the resistance of the wire may end up between the P and AP state. It is assumed that the magnetization of the free electrode is reversed only in a certain part of the wire and that two domain walls exist in the wire after the current pulse. MR measurements (not shown) indicate that the domain walls are trapped at edge roughnesses leading to a step in the switching characteristics.

The switching can be explained in the following way (see schematics in Fig. 3(b)): The current through the copper layer generates an Oersted field which is approximately transverse to the wire (see also [14]) and points in opposite directions for the adjacent ferromagnetic layers. During a current pulse the magnetizations of the free (large angle) and the hard (small angle) electrode are thus laterally rotated out of the wires axis [15] in opposite directions creating massive stray fields. Even strong transversal magnetic fields cannot rotate the magnetization of a single layer by more than 90° with respect to the wire axis, but additional dipolar coupling due to the induced stray fields in PSV structures can drag the free layer into a state which is antiparallel to the hard layer and thus rotate by more than 90°. When the Oersted field vanishes, the system relaxes into an AP state where the hard layer's magnetization is still pointing in its initial direction along the wire. Applying a current to a wire in the AP state in zero magnetic field yields no switching effect.

Others [16] investigated the current-induced dynamics in PSV stripes by using time-resolved x-ray photoemission electron microscopy. Their investigations show a large tilt (but no switching) of the magnetization of the free electrode during the pulses supporting our model.

We have investigated several factors which affect the critical current density $j_{crit}$, such as free layer thickness, width of the PSV wire and small external magnetic fields. Figure 4(a) (insert) shows that $j_{crit}$ increases strongly with the free layer thickness. For an increase of the free layer thickness from 3 to 5 nm, $j_{crit}$ increases by a factor of 4. The dependence of the critical current density on the wire width (see Fig. 4(a)) is less pronounced, but $j_{crit}$ is still increased by almost 50 % when the wire width is decreased from 500 nm to 200 nm. Small external magnetic fields $\mu_0 H_{ext}$ along the wire affect the current density depending on the direction of the field (see Fig. 4(b)). For fields that favor the P state the critical current density is increased, for reversed fields it is decreased. For "large" external fields the switching can only be shown for fields favoring the AP state (reduction of $j_{crit}$).

Obviously a number of side effects must be considered which may lead to similar behavior. Electrical contacts formed by narrow stripes can induce domain walls in the wire due to local magnetic fields [17]. To exclude domain wall induction we use large contacts inducing low Oersted fields that cover the full end of the wire as standard geometry for our experiments (see Fig. 1(a)). Passing current through a nanowire is usually accompanied by Joule heating which can be substantial [7], [18]. The heating can reduce both, magnetization and coercive field, and, once the Curie temperature ($T_C$) is exceeded, the structure can relax into an AP state when the temperature drops again below $T_C$ [19]. We calibrate the temperature of a nanowire by measuring the resistance of the externally heated sample and comparing to resistance values (real time measurements using an oscilloscope) during a critical current pulse. During a current pulse with critical current density in zero field the resistance of a particular wire is increased by 4 % corresponding to a temperature increase of not more than 100 °C resulting in a temperature well below $T_C$ of the magnetic layers. To ascertain this fact we have also measured the magneto resistance at T = 200 °C obtaining still half of the room temperature value. For a detailed analysis of the temperature distribution in the nanowire, simulations of both the wire and the substrate (see for example [20]) have to be carried out.

Vertical temperature gradients are also considered. Although the temperature is not close to the Curie temperature the different thermal coupling at top and bottom interface can cause a perpendicular thermal gradient. Theoretically this gradient can cause a switching by thermally induced spin transfer torque [21], [22]. To investigate the possibility of thermal spin transfer torque induced switching, samples with reversed layer order are also investigated. Depending on the layer sequence a thermal gradient should either point from the fixed to the free layer or from the free to the fixed layer and thus either favor a P or an AP alignment. However, in both structures only switching from the P to the AP state is observed. Another coupling mechanism known to be critical in short stripes is the dipolar coupling via the ends of the wire [14]. We can rule out this mechanism using stripes with numerous contacts at different positions along the wire (see Fig. 1(b)) and L-shaped wires (see Fig. 1(c)). By sending a current through a part of the wire which is remote from both ends (contacts B, C in Fig. 1(b)) we are able to switch the magnetization of just that segment of the wire while other parts of the wire remain in the parallel state. To exclude a moving domain wall that might be present in the wire after the reset we use the same geometry and compare the following two experiments. First we switch the AB part of the wire leaving a domain wall beneath contact B. Subsequently we send current pulses through the whole structure (AD) and determine the current density necessary to switch the BC part. As a second step we repeat the experiment without a domain wall beneath contact B. Both experiments yield comparable critical currents indicating that the domain wall beneath contact B cannot be moved by our current pulses and does furthermore not influence the critical current density.

The same switching phenomenon is equally observed in $Co_{20}Fe_{60}B_{20}$ free electrodes in zero magnetic field (see Fig. 3(a)). Critical current densities of $(2.3 \pm 0.1) \cdot 10^{12}$ A/m$^2$ for full magnetic reversal of the free layer were measured for 150 nm wide stripes with a free layer thickness of 3 nm. In 300 nm wide stripes with 8 nm Cu we find critical current densities as low as $(0.9 \pm 0.2) \cdot 10^{12}$ A/m$^2$.

The effect that we observe also does not rely on a homogeneous magnetization of the hard magnetic layer. Even if the hard magnetic layer shows a multi domain state, the free layer will show the exact antiparallel replica including all domain walls after the current pulse has been applied. This can be demonstrated using L-shaped wires (see Fig. 1(c)). Using suitable external field sequences one can prepare different magnetic states in the L-wire [7]. For states without a domain wall at the bending the current pulse results in a continuously magnetized antiparallel state, uninterrupted by any domain wall. If, however, the parallel (starting) state includes a domain between the two legs of the L-structure (head-to-head or tail-to-tail configuration) the current pulse converts the free layer from head-to-head starting state to a tail-to-tail end state and vice versa.

To confirm our model we carry out micromagnetic simulations using Mumax 2 [12]. We simulate 4096 nm long tapered stripes (see Fig. 5) with 128 nm width and a slight edge roughness. The simulated layer sequence is NiFe / Cu / CoFe. The wire is oriented along the x-axis. The y-axis lies parallel to the width of the wire and the z-axis perpendicular to the layer plane. We use a saturation magnetization of 1400 kA/m for CoFe (800 kA/m for NiFe [23]). The Gilbert damping is set to 0.006 for the magnetic layers. The whole structure is divided into $4 \times 4 \times 1$ nm$^3$ cells, 1 nm in the z-direction. Two contributions to the magnetic field are considered, namely an external magnetic field and the Oersted field. The Oersted field is applied to the central part of the wire (between the contact pads), the external magnetic field is applied to the complete structure. Starting with a P state along +x-direction we apply the Oersted field after a 2 ns relaxation period. The Oersted field increases linearly in time for 3 ns (slow field rise to exclude precessional dynamics) to its maximum value. The field remains constant for 7 ns, and is then linearly decreased to zero. The results are analyzed after another 2 ns relaxation period. During application of the Oersted field the magnetizations of the NiFe and the CoFe layers are rotated in the layer plane. The rotation angle differs significantly for the different materials. For typical fields the magnetization of the NiFe (CoFe) layer is tilted by an angle near 90° (10°) in the center of the wire. Depending on the magnitude of the Oersted field the tilt angle of the NiFe magnetization exceeds 90°. With the Oersted field removed, this state relaxes not to the initial P state but to the AP orientation.

In our simulations switching only takes place for thickness ratios of the magnetic layers in a certain range. For a copper thickness of 5 nm switching is observed for NiFe/CoFe ratios between 2/6 and 2/10. For ratios smaller than 2/11 the magnetization of the NiFe is not stable in zero field meaning that a magnetization reversal starting from the ends of the wire occurs without a transversal field. No switching occurs in simulations with thickness ratios larger than 2/5 contradictory to our experiments where switching could be demonstrated at least for ratios between 3/10 and 5/10. The structure width also influences the critical field. In the range between 64 nm and 160 nm wide structures, the critical field reduces in simulations for a 2/10 thickness ratio and 5 nm Cu thickness from 29 mT to 19 mT by roughly 65 %. Increasing the widths of nanostripes with a thickness ratio of 3/10 in the experiment from 200 to 500 nm reduces the critical current in good agreement with the simulations by about 60 % (Fig. 3 (a)). The experimental Oersted field in our structures is estimated by calculating the transversal field component of a current density of an infinitely long rectangular conductor with the lateral dimensions of the copper layer. A current density of $1.5 \cdot 10^{12}$ A/m$^2$ through a $200 \times 5$ nm$^2$ wire generates a transversal field of 4.5 mT near its surface. Comparing the critical fields of simulations and experiments shows that the simulated fields are about a factor of 4 higher than observed in the experiment. One explanation for this discrepancy is the fact that our simulations are done at 0 K and thermal fluctuations are likely to reduce the critical field.

Although our simulations show that the Oersted fields may be responsible for the switching, another possible mechanism which cannot be ruled out in these experiments may contribute. For a parallel orientation of the magnetizations in the PSV structure electrons flowing in the copper layer are expected to have either a high or low scatter probability at each Cu / FM interface. As a consequence the backscattered electrons are spin polarized opposite to the FM. This corresponds to an explanation of GMR based on a Fuchs-Sondheimer description of transport in thin metal layers [24], [25]. These spin polarized electrons may exert a torque on the magnetization of the free magnetic layer when they are scattered at the Cu / free FM interface leading to a spin transfer torque as has been observed in vertical nanostructures. For symmetry reasons this in-plane spin transfer torque has an effect fundamentally different from the one in vertical structures. In vertical GMR structures the STT either favors the parallel or the antiparallel state depending on the current direction. In a lateral structure, however, the scattering of electrons between the two layers always favors the antiparallel alignment and thus supports the Oersted field based switching. Since the critical current density decreases with increasing structure widths (see Fig. 3(b)) the influence of the Oerstedfield is supposed to dominate, but most likely the observed effect is merely a combination of the two mechanisms. It may be possible to further distinguish between the two by carrying out experiments with exceedingly thick Cu interlayers, however, at the cost of a minute magneto resistance and maybe even a reduced magnetic coupling between the two magnetic layers.

In conclusion, we demonstrate that PSV nanostripes with very thin free electrodes can be switched from the P to the AP state by the current-induced Oersted field. The critical current density in zero magnetic field for NiFe (CoFeB) electrodes is found to be in the range of $1.5 \cdot 10^{12}$ A/m$^2$ ($2.3 \cdot 10^{12}$ A/m$^2$) for 200 (150) nm wide stripes. Even small external magnetic fields can reduce the current density to $5 \cdot 10^{11}$ A/m$^2$ in our experiment (NiFe electrode). As a consequence one should keep in mind that the influence of the Oersted fields might be dramatic for in-plane transport experiments with very thin free layers in PSV structures and can even lead to an unwanted reversal of the magnetization. Especially when investigating domain walls special care is needed to separate Oersted field from spin torque effects.

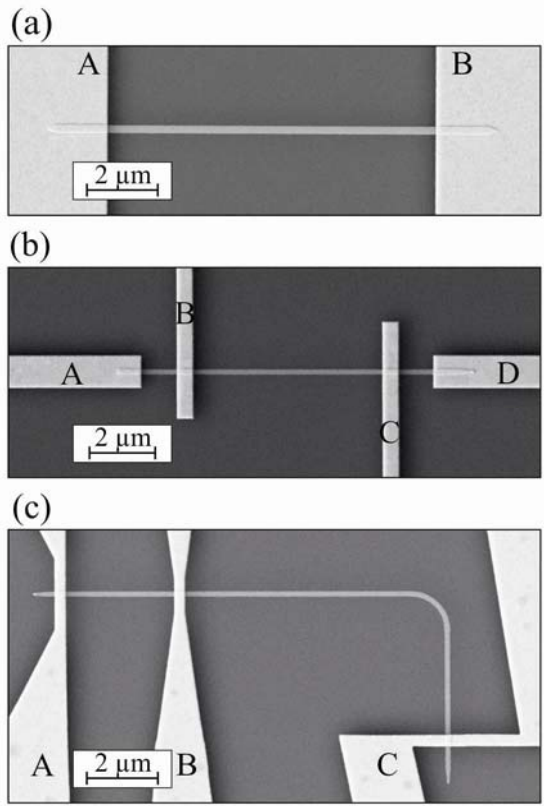

*Figure 1: Overview (SEM micrographs) of nanowires with different contact geometries used in this study. (a) standard geometry with large contact pads. Example shows a 200 nm wide, 10 µm long GMR stripe. (b) Geometry used for 4-probe measurements and experiments where only parts of the wire are switched. Example shows a 150 nm wide, 9 µm long stripe. (c) L-shaped wire with 3 contacts.*

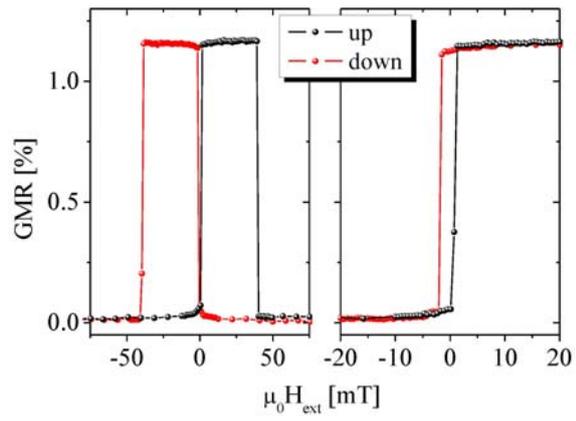

*Figure 2: MR measurements (left: major loop measurement, right: minor loop measurement) of a tapered GMR PSV nanowire (200 nm wide, 10 µm long, 3 nm NiFe free layer and 5 nm Cu spacer).*

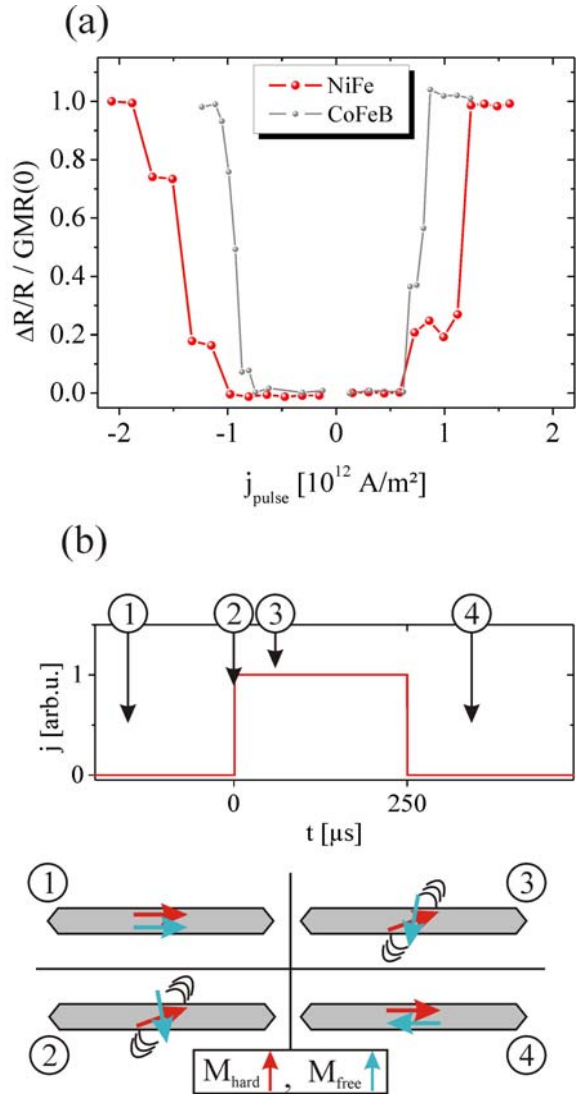

*Figure 3: Zero magnetic field current-induced magnetization reversal of PSV nanowires after reset at $-150$ mT. Results for structures with free electrodes from NiFe (200 nm wire) and CoFeB (300 nm wire with 8 nm Cu layer) are shown. (b) Schematics showing a current pulse and corresponding magnetization orientations in the nanowire.*

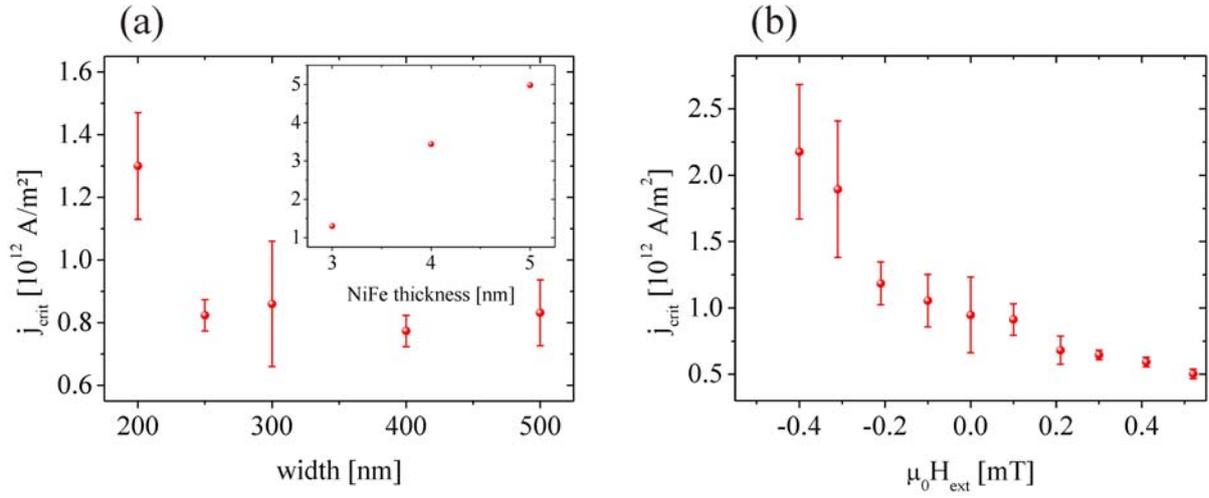

*Figure 4*: *Critical current density as a fuction of structure width (a), free layer thickness (insert) and external magnetic field* $\mu_0H_{ext}$ *(b). (a) shows a moderate increase in* $j_{crit}$ *for narrow stripes, whereas* $j_{crit}$ *increases strongly with the free layer thickness for* 200 nm *wide stripes. (b) shows the variation of the critical current density in small external, inplane magnetic field applied in wire direction. All measurements represent structures with standard geometry.*

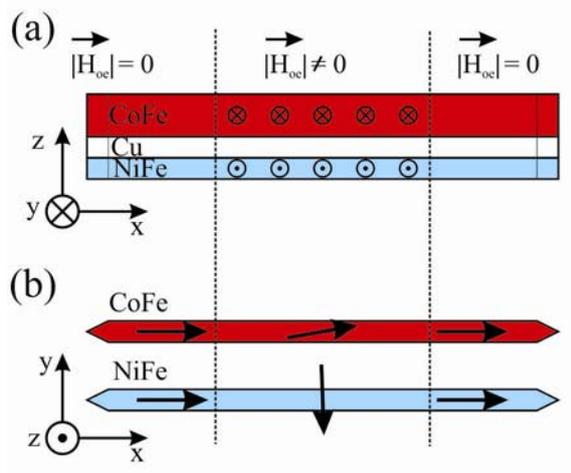

*Figure 5: Sketch of the simulated structure, where (a) shows a side view. The x-axis points along the wire, the z-direction perpendicular to the layer planes. Between the contact pads the Oersted field of the current through the Cu layer is considered as opposite transversal magnetic fields in the adjacent magnetic layers. Due to the Oersted field the magnetizations of the electrodes rotate in different directions, (b) shows top view. The rotational angle of the free layer is much larger than that of the hard electrode.*